\documentclass[11pt,twoside]{article}
\usepackage{asp2004}
\usepackage{psfig}
\usepackage{epsf}
\usepackage{graphics}
\usepackage{lscape}
\def\edcomment#1{\iffalse\marginpar{\raggedright\sl#1\/}\else\relax\fi}
\reversemarginpar

\begin{document}
\title{Extra-planar H\,I in the Inner Milky Way}
\author{Felix J. Lockman}
\affil{National Radio Astronomy Observatory,\\ P.O. Box 2, Green Bank WV
24944, USA}
\author{Yurii Pidopryhora}
\affil{Ohio University, and National Radio Astronomy Observatory,\\
P.O. Box 2, Green Bank WV 24944, USA}

\begin{abstract}

The extra-planar H\,I in the inner parts of the Milky Way
has been discovered to contain numerous cloud-like structures
when observed in the 21cm line with the Green Bank Telescope.
These halo clouds have motions consistent with Galactic rotation and
do not seem to be related to the classic high-velocity clouds.
They are found to distances $>$1 kpc from the plane and can
contain hundreds of $M_{\sun}$ of H\,I. Spectra of many of the halo clouds
show evidence of coexisting cool and warm H\,I phases. A preliminary
high-resolution study of one of the clouds suggests that it consists
of a diffuse envelope and a few dense cores, with a peak $N_{H\,I}$ reaching
$4 \times 10^{20} $ cm$^{-2}$. The clouds are often organized into larger
structures, one example of which was discovered near $\ell=35\deg$
rising higher than 2 kpc above the Galactic plane. New observations should
answer some fundamental questions about the nature of these clouds.

\end{abstract}
\thispagestyle{plain}

\section{Introduction}

The Milky Way provides an opportunity to study extra-planar gas
 in great detail, yet the view is often confused and uncertain:
at high latitudes where spectra can be simple, there is rarely any
way to estimate a distance to the gas being measured,  while at
low latitudes, where Galactic rotation can  separate gas kinematically,
 spectra are blended and confused.
Neutral gas in significant amounts has been found far from the Galactic plane,
and measurements in optical, radio, and UV absorption lines
have provided detailed information on its physical properties,
but the data do not form a consistent picture, do not
 reveal the structure of the material, and leave us with
questions about how to generalize from one observation to the next
 \citep{munchzirin,savagedeboer,l84,lhs,savage90,
spitzfitz93,albert,sembsav,spitzfitz97,kalberla98,howk_sf}.

Our understanding of the lower halo
 changed dramatically upon commissioning of the new
 Green Bank Telescope (GBT) \citep{fjlspie98, jewellspie} whose $9\arcmin$
angular resolution and superb  sensitivity to low surface brightness
emission revealed structures in Galactic H\,I which were only
hinted at in earlier 21 cm data.
We now know that much of the extra-planar H\,I in the inner parts of the Milky
Way is organized into discrete objects, some of which look like
archetypical diffuse interstellar
 clouds \citep{fjl2002, fjl2004}.  The distance to many of the clouds
can be estimated with some accuracy, allowing us to derive
 quantitative information on their temperature, density, and mass.
  We are  now in a
new era in the study of the Galactic halo: it may come to
be the model for the disk-halo interface in other systems as well.

\section{H\,I Clouds in the Halo}

Figure 1 shows the 21cm H\,I emission measured with the GBT along
a cut through the Galactic plane at $\ell =24\fdg4$.
The fainter emission displayed in the gray scale has
a distinctive cloudy structure.   The Figure illustrates
several key points about extra-planar H\,I in the inner Milky Way:

\begin{figure}[!ht]
\plotfiddle{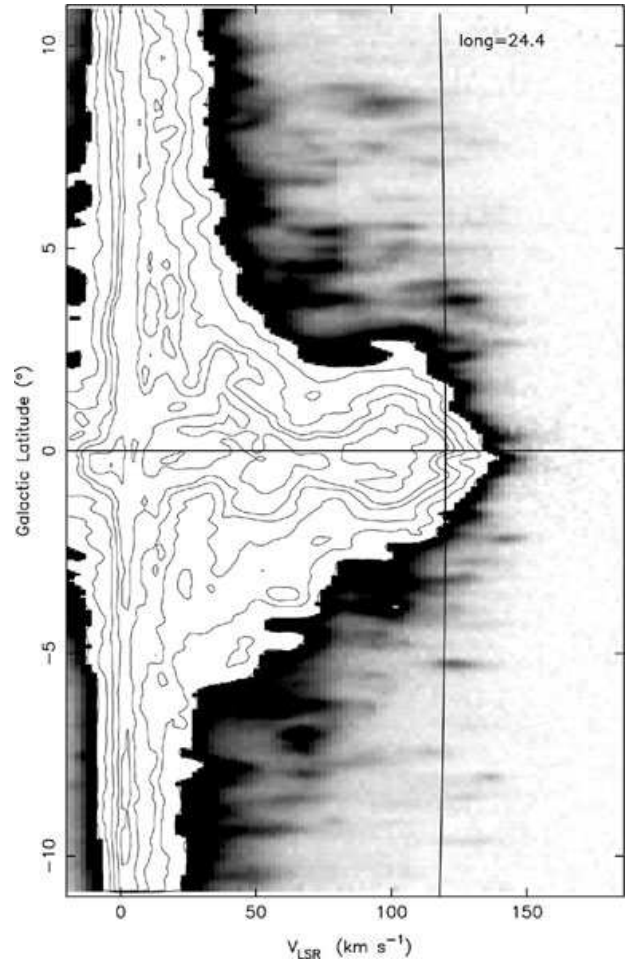}{4.5in}{0.0}{100}{100}{-310}{-270}
\vspace{.75in} \caption{Neutral Hydrogen emission observed with
the Green Bank Telescope at $9\arcmin$ angular resolution along a
cut through the Galactic plane at longitude $24\fdg4$.  The bright
emission is marked with contours
 at 5, 10, 20, 35, 50 and 80 K, while the emission between
zero and 5 K is displayed on a gray scale.
  In this direction Galactic rotation
produces LSR velocities up to $\approx 120\ cos(b)$ km s$^{-1}$
at the tangent point (the vertical line), which lies
 at a distance $r=R_0\ cos(\ell) = 7.7 $ kpc.}
\end{figure}

\begin{itemize}
\item  H\,I emission is seen to at least $10\deg$ from the
 plane over the same velocity range as is found in the disk.
  This extra-planar or  halo H\,I is  kinematically
well behaved and its velocity arises mainly from Galactic rotation.
\item Much of the H\,I away from the Galactic plane seems concentrated
into discrete structures which are isolated in position
and velocity: clouds.

\item Clouds and cloud-like structures are seen even at low latitudes, but
they are usually  heavily blended, and difficult to distinguish.

\item Many halo clouds have a  narrow linewidth indicating
that they contain cool gas.

\end{itemize}

Another aspect of Galactic extra-planar H\,I, though not so apparent
in Fig.~1, is that clouds
are sometimes found in groups connected by a common  diffuse
envelope or filament, and occasionally are
 organized into  larger  structures on scales $\sim 1$ kpc.
The remainder of this paper will expand on these points and
their implications.

\section{Kinematics}
 It is clear from Fig.~1 that H\,I above and below the plane shares the
same kinematics as material in the disk and thus its motion
is predominantly due to Galactic rotation.
The gas considered here  is therefore not related to the classic
`high-velocity' clouds whose kinematics have a large anomalous
motion \citep{wvw97}.  This fact makes it possible  to determine  reasonably
accurate distances to some of the emission features.

 At the longitude of Figure 1, Galactic rotation
produces $V_{LSR} \la 120$ km~s$^{-1}$
(this is established from observations of giant molecular
clouds in the disk, e.g., \citet{clemens}).  In the
first Galactic quadrant,  the rotation projects  a maximum LSR velocity
$V_t$ at the tangent point, whose distance  is known from simple
trigonometry: $r = R_0 cos(\ell) = 7.7 $ kpc.  Thus  the gas
in Fig.~1 at $V_{LSR} \approx 120 $ km s$^{-1}$ (marked with the
vertical line) probably lies close to the tangent point where
 each degree in latitude corresponds to a displacement of about
 135 pc from the Galactic plane.

There may be a component of halo H\,I
whose velocity lags behind the rotation of the disk, as is
observed in a number of systems described elsewhere in these
Proceedings.  Such  lagging halo gas would lie at $V_{LSR}<V_t$
and be confused with other emission.  The existence of a lagging
halo is the subject of another work -- here we concentrate on
gas whose kinematics follows Galactic rotation, and specifically
on H\,I which is likely to be near the tangent point where
distance uncertainties are the smallest.

\begin{figure}[!ht]
\plotfiddle{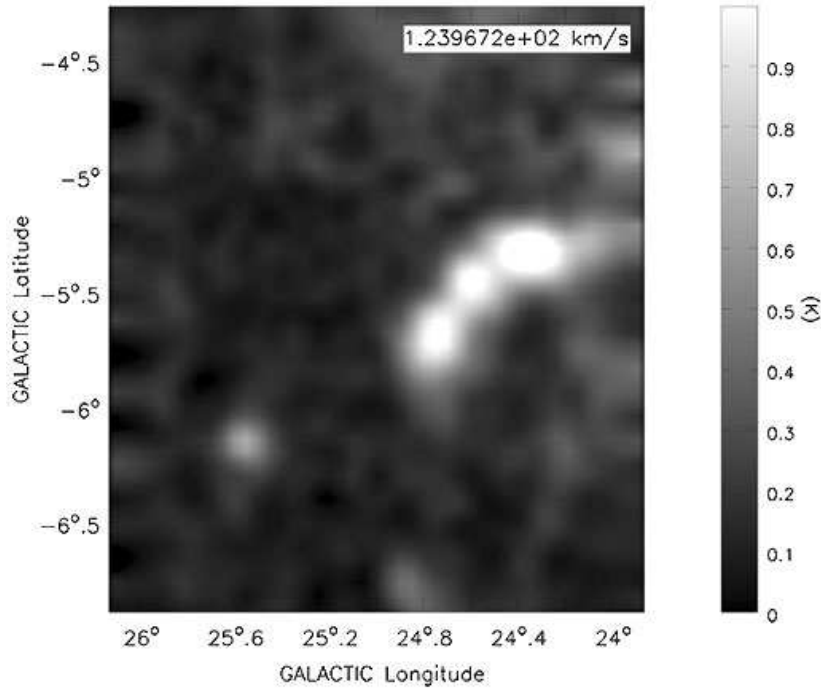}{4.0in}{-0.0}{100}{100}{-300}{-270}
\vspace{.0in} \caption{GBT observations of 21cm HI emission from
several halo HI clouds   at $V_{LSR}=124$ km s$^{-1}$ in a field
around $l,b$=25\deg-$5\fdg5$. This velocity is slightly beyond
that due to Galactic rotation, so the clouds are probably quite
near the tangent point at a distance $z = $ --700 pc from the
Galactic plane.}
\end{figure}

\section{Clouds}
We describe the extra-planar H\,I  as `cloudy' because
that is exactly how it looks.  Many clouds have a spheroidal
shape or are resolved into quasi-spherical objects.
 Figure 2  shows a group of three clouds near $24\fdg$6-$5\fdg5$
and a fourth fainter one at $25\fdg6$-$6\fdg2$.
The cloud at $24\fdg4$-$5\fdg3$  appears also in the velocity-latitude
cut of Figure 1.   The chain of three clouds
is about 100 pc long and lies  $700$ pc below the Galactic plane.  The
individual clouds have identical LSR velocities to within 2~km~s$^{-1}$.  The
total H\,I mass of the structure is  1350~M$_{\sun}$, about
70\% of which is in the clouds and the remainder in a diffuse envelope
which surrounds them.  Individual clouds have H\,I masses of 410, 170
and 350~M$_{\sun}$ right to left.
The cloud at $25\fdg6$-$6\fdg2$ seems unrelated to the
others.   It contains 70~M$_{\sun}$ of hydrogen and is unresolved
 by the GBT.  It must have a diameter  $\leq 20$ pc.

Clouds from a sample of 40 detected with the GBT at longitude $29\deg$ at
 a median distance from the plane of --940 pc
have a median H\,I mass of 50~M$_{\sun}$
and a median diameter of a few tens of pc \citep{fjl2002}.  Some
contain lines so narrow that their kinetic temperature must
be $<1000 $ K.  The internal structure of halo clouds has
recently been measured at higher angular resolution, and
preliminary results are discussed in \S5.

Some halo clouds are dense, cool, concentrations embedded in
 filamentary structures of broad-line H\,I which contain most
of the mass \citep{fjl2004}.  It is  an open question as
to the fraction of the total halo H\,I concentrated in discrete
structures, though it appears to be the majority in some regions.
Considerable progress on this issue can be expected in the
next year as observations of increased sensitivity are made.

\section{Internal Structure}

   The spectra of many halo clouds  consist
of at least two components, one broad and one narrow, with
typical linewidths (FWHM) of 20 and 6 km s$^{-1}$, respectively.
  This suggests that neutral hydrogen in the clouds is in two phases,
one warm and one cool,
a structure which is seen in some high-velocity clouds, and which is
expected theoretically at certain  pressures \citep{ferrara94,wolfire95}.
Systematic differences in the internal structure of
halo clouds with location in the Galaxy and distance from the
plane  are now being observed.   The clouds may be unique
 probes of  physical conditions in the halo.

\begin{figure}[!ht]
\plotfiddle{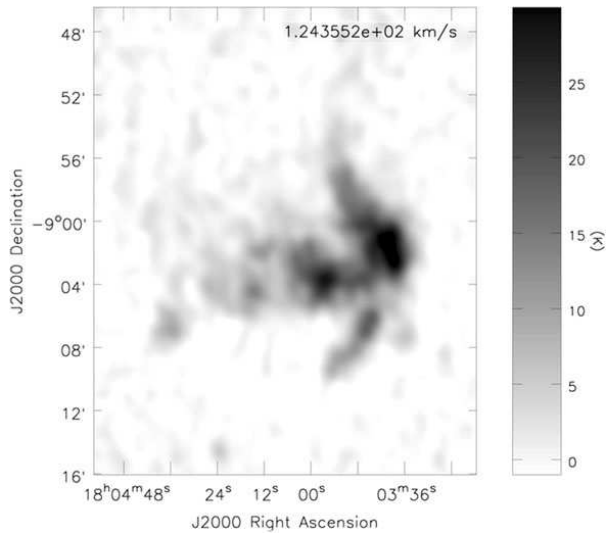}{2in}{-0.0}{80}{80}{-255.0}{-270.0}
\vspace{.75in} \caption{Preliminary VLA 21cm H\,I channel map at
$1\arcmin$ angular resolution of the halo cloud at $\ell,b =
19\fdg4$+6$\fdg3$
 \citep{yurii}.
This map is in equatorial coordinates: the Galactic plane is
toward the lower left side of the Figure.  The cloud
 lies 2.8 kpc from the Galactic center and 900 pc above the Galactic plane.
}
\end{figure}

    Very recently several halo clouds have  been observed
with the VLA in the 21cm line  at an angular resolution of about
$1\arcmin$, which corresponds to a linear resolution of a few
pc.  The data are not yet fully analyzed, but a preliminary channel
map for one cloud is shown in Figure 3.  The brightest part
of this cloud has a peak $N_{HI} = 4 \times 10^{20} $ cm$^{-2}$
and a core with a linear size of about 7 pc, giving an average
density $\langle n \rangle = 20$ cm$^{-3}$.  The line at this position has
a FWHM $= 6$ km s$^{-1}$, limiting the kinetic temperature to
$T \leq 800$ K.  These values can be used to estimate a thermal
pressure $P/k = n/T \leq 16000$ cm$^{-3}$ pc.  A preliminary
accounting suggests that perhaps half of the 800 $M_{\sun}$
of H\,I in this cloud is contained in its denser central structure, with
the remainder distributed more broadly. The three most compact
cores hold about 110, 70 and 10 $M_{\sun}$.

    The typical halo cloud studied so far has a peak
$N_{HI}$  below the $5 \times 10^{20}$~cm$^{-2}$
  required  to allow formation of molecular hydrogen \citep{bsd78},
but there may be denser regions in some clouds and
a significant molecular component.

\begin{figure}[!ht]
\plotfiddle{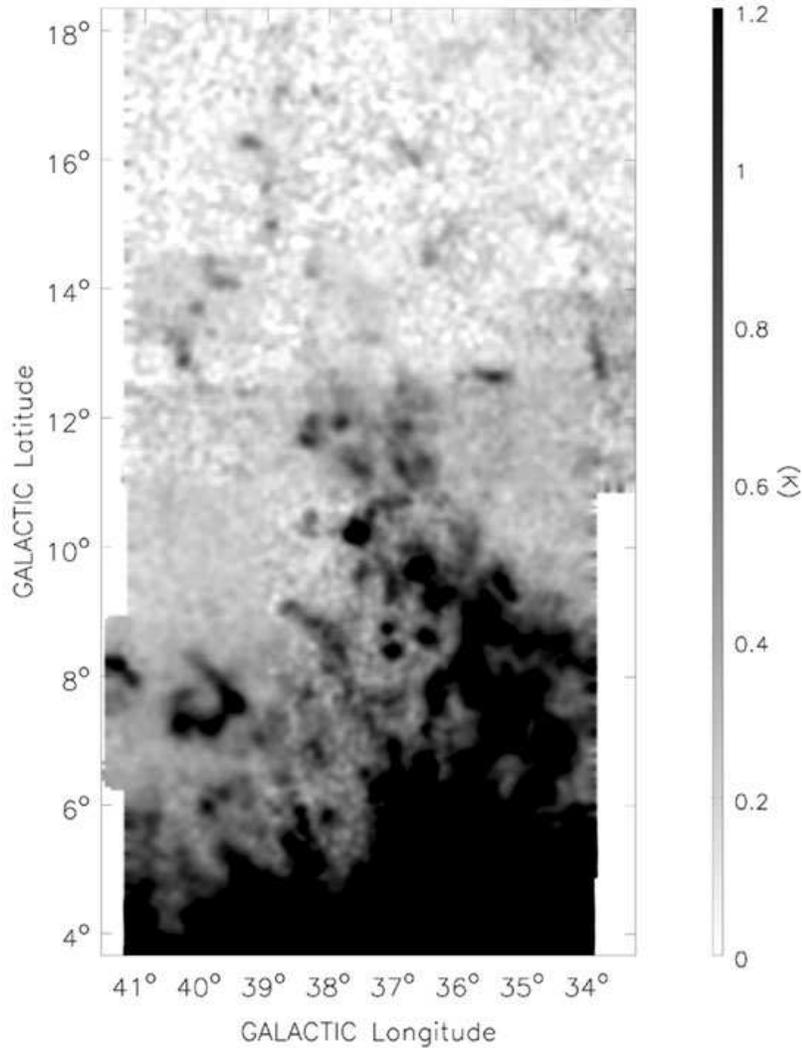}{5.0in}{-0.0}{92}{92}{-280.0}{-205.0}
\vspace{.5in} \caption{A large  H\,I structure rising out of the
Galactic plane near longitude $35\deg$.  This image shows GBT data
averaged over 3 km s$^{-1}$ around +79 km s$^{-1}$. At the
location of this `whisker' one degree of latitude corresponds to
about 120 pc, so the structure extends
 more than 2 kpc above the plane.}
\end{figure}

\section{Large-Scale Organized Structures}

While there are many H\,I clouds in the halo which appear
as discrete isolated objects, near longitude $35\deg$
there is an  enormous coherent structure of gas rising from
the Galactic plane to heights of more than 2 kpc (Figure 4).
This `whisker' of gas contains more than $10^5 \, M_{\sun}$
of H\,I.  In its lower regions it appears more like a
`turbulent network' \citep{chappellscalo} than a set of
discrete clouds, but by $b \ga 10\deg$ ($z \ga 1200$ pc)
is appears to break up into clusters of  clouds. Some of
the `whisker' clouds are seen even to latitudes higher than $20\deg$
($z > 2$ kpc) --- the record height for the halo clouds discovered so far.
A preliminary study indicates that while `mid-latitude' ($b \approx  10\deg$)
`whisker' clouds share the two-component spectral structure
of most halo clouds, high-latitude clouds have only
the broad component (FWHM $\ga$ 20 km s$^{-1}$). This may indicate that
only a warm ISM phase survives at $z \ga$ 2\,kpc.

Study of the `whisker' feature is just beginning; it may be a local
counterpart to the vertical dust lanes seen in
galaxies like NGC 891  \citep{howkn891}.

\section{The Origin of a Cloudy H\,I Halo}

From the GBT observations of the inner Galaxy we get the definite
sense that nature likes to gather hydrogen into clouds. There is
even evidence that the population of halo clouds extends right
down into the plane \citep{lockmanstil, stil05}. Where did the
clouds come from and what keeps them together?

The clouds do not appear to be self-gravitating,  yet their
dynamical time (size/linewidth) is only a few million years.
 If the clouds
are long-lived they must be confined by an external
medium, presumably the Galaxy's hot halo
\citep{spitzer56,munchzirin}. Even if they are in pressure equilibrium,
they are denser than their surroundings by orders of magnitude and
should be dropping toward the plane like stones, with a
 free-fall time of a few $10^7$ years.  Are they constantly
reforming from condensing hot gas in a galactic `fountain' \citep{shapiro,
bregman80,houckbregman}?
Or are they fragments of structures launched upward from below, as the
morphology of the `whisker' suggests (cf. the models of \citet{deAvillez00})?

Hints of core-envelope structure combined with indications of mixed
ISM phase coexistence, and even the size and mass range of the clouds
are reminiscent of theoretical models like those of \citet{mo},  but we do
not know if the halo clouds are evaporating or accreting mass, we do
not even know if they are rising, falling, or steady in place.
Nor do we know the relationship, if any, between the halo H\,I
and the Galaxy's thick ionized layer \citep{reynolds90}.
 \citet{dwark} has recently proposed that the halo clouds
are the distant analog of the `high velocity-dispersion'
cloud population observed in local optical and 21cm absorption lines
 \citep{routlyspitz,rajagopal}.  If the local clouds could be
detected in 21cm emission maps, a direct comparison of their properties
with the halo clouds might settle this issue.

Studies of H\,I above the Galactic plane in the inner Galaxy
have thus far measured only co-rotating gas \citep{l84, dl90}.
If there is a significant lagging
component,  estimates of the mass and extent of the
neutral halo are too  low.  Questions far outnumber answers
today, but there are  good prospects for rapid progress
as new observations continue to reveal more details of
the Milky Way's fascinating extra-planar gas.

\end{document}